# Gas transport in partially-saturated sand packs

Behzad Ghanbarian[1], Shoichiro Hamamoto[2], Ken Kawamoto[3], Toshihiro Sakaki[4], Per Moldrup[5], Taku Nishimura[2], and Toshiko Komatsu[3]

[1] Porous Media Research Lab, Department of Geology, Kansas State University, Manhattan 66506 KS, USA

[2] Graduate School of Agricultural and Life Sciences, The University of Tokyo, 1-1-1, Yayoi, Bunkyoku, Tokyo 113-8657, Japan

[3] Graduate School of Science and Engineering, Saitama University, 255, Shimo-okubo, Sakuraku, Saitama, 338-8570, Japan

[4] Department of Civil and Earth Resources Engineering, Kyoto University, Kyoto, 615-8540 Japan

[5] Department of Civil Engineering, Aalborg University, Thomas Manns Vej 23, DK-9220 Aalborg, Denmark

* Corresponding author's email address: ghanbarian@ksu.edu

**Abstract**

Understanding gas transport in porous media and its mechanism has broad applications in various research areas, such as carbon sequestration in deep saline aquifers and gas explorations in reservoir rocks. Gas transport is mainly controlled by pore space geometrical and morphological characteristics. In this study, we apply a physically-based model developed using concepts from percolation theory (PT) and the effective-medium approximation (EMA) to better understand diffusion and permeability of gas in packings of angular and rounded sand grains as well as glass beads. Two average sizes of grain i.e., 0.3 and 0.5 mm were used to pack sands in a column of 6 cm height and 4.9 cm diameter so that the total porosity of all packs was near 0.4. Water content, gas-filled porosity (also known as gas content), gas diffusion, and gas permeability were measured at different capillary pressures. The X-ray computed tomography method and the 3DMA-Rock software package were applied to determine the average pore coordination number $z$. Results showed that both saturation-dependent diffusion and permeability of gas showed almost linear behavior at higher gas-filled porosities, while deviated substantially from linear scaling at lower gas saturations. Comparing the theory with the diffusion and permeability experiments showed that the determined value of $z$ ranged between 2.8 and 5.3, not greatly different from X-ray computed tomography results. The obtained results clearly indicate that the effect of the pore-throat size distribution on gas diffusion and permeability was minimal in these sand and glass bead packs.

# 1. Introduction

Understanding mechanisms controlling gas transport in porous materials has wide applications, particularly in carbon sequestration in deep aquifers and natural gas explorations in geological formations. Both geometrical and morphological properties of pore space influence gas flow and transport. Accordingly, numerous empirical, quasi-physical, and theoretical models were developed to address effects of different factors e.g., pore size distribution, porosity, connectivity, and tortuosity on gas diffusion and gas permeability in porous media. In what follows, we briefly review several saturation-dependent models proposed in the literature to study diffusion and permeability of gas and discuss their applications.

## 1.1 Empirical models

Literature on gas transport models is vast and extensive. Various empirical models were proposed to characterize saturation-dependent gas diffusion and permeability in porous media. For example, Buckingham (Buckingham, 1904) suggested a power-law model to relate gas diffusion to gas-filled porosity. Years later, Penman (Penman, 1940), however, proposed a linear function for saturation-dependent gas diffusion in a wide variety of porous media. Another notable empirical model is the following relationship by Troeh et al. (Troeh et al., 1982):

$$\frac{D(\varepsilon)}{D_0} = \left[\frac{\varepsilon - u}{1 - u}\right]^v \tag{1}$$

where $D(\varepsilon)$ is the gas diffusion coefficient in porous medium, $D_0$ is the gas diffusion coefficient in free space, $\varepsilon$ is the gas-filled porosity (also known as gas content), and $u$ and $v$ are empirical constant coefficients.

Gardner and Mayhugh (Gardner and Mayhugh, 1958) proposed an empirical relationship to describe relative permeability. The Gardner model which has an exponential form has been frequently used to determine relative permeability in porous media. However, their model parameters are priori unknown. Accordingly, they should be determined by directly fitting Gardner's model to experimentally measured observations. In what follows, we discuss quasi-physical and theoretical models whose most parameters, if not all, are physically meaningful and can be determined from geometrical and/or topological properties of porous media.

## 1.2. Theoretical and quasi-physical models

In addition to empirical methods, theoretical and quasi-physical models were applied to study gas transport in porous media. Numerous quasi-physical models were developed based on a bundle of capillary tubes approach see e.g., (Burdine, 1953; Moldrup et al., 1999; Xu and Yu, 2008), while other theoretical models (Hunt et al., 2014; Sahimi, 2011, 1994) were proposed using percolation theory (PT) and the effective-medium approximation (EMA). Here we mainly review those theoretical models from PT and the EMA.

Universal scaling law from percolation theory, a power law in the gas-filled porosity (less a critical gas-filled porosity) with an exponent of 2, was successfully used to describe the saturation dependence of gas diffusion (B. Ghanbarian et al., 2015a; Ghanbarian and Hunt, 2014; Hamamoto et al., 2010) and gas permeability (Ghanbarian-Alavijeh and Hunt, 2012b; Ghanbarian et al., 2015b; Hunt et al., 2014; Hunt, 2005) in natural porous media. For a recent comprehensive review, see (Hunt and Sahimi, 2017).

By combining the universal scaling from percolation theory with the linear scaling from the effective-medium approximation, one obtains the following model for diffusion in a *percolating lattice* (Hunt et al., 2014):

$$\frac{D(p)}{D_0} = \begin{cases} \frac{p_x - 2/z}{1 - 2/z}\left[\frac{p - p_c}{p_x - p_c}\right]^2, & p_c \leq p \leq p_x \\ \frac{p - 2/z}{1 - 2/z}, & p_x \leq p \leq 1 \end{cases} \quad (2)$$

where $D(p)$ is diffusion coefficient in the lattice, $D_0$ is diffusion coefficient in free space, $z$ is the average coordination number, $p$ is the occupation probability ($0 \leq p \leq 1$), $p_c$ is the percolation threshold, and $p_x$ is the crossover probability at which behavior transitions from percolation scaling to the EMA scaling. The significance of a crossover between the EMA and percolation description of the conductivity was previously pointed out by Kirkpatrick (1973) and Sahimi et al. (1983).

Ghanbarian and Hunt (2014) and Hunt et al. (2014) set $2/z = p_c$ (percolation threshold) in Eq. (2) and found well agreement with respectively 66 gas and 106 solute diffusion experiments from the literature. More recently, Ghanbarian et al. (2015a) incorporated the effect of coordination number and showed that the crossover point in their model was very clear in the lattice-Boltzmann simulations of gas and solute diffusion in mono-size packings of overlapping or non-overlapping spheres. Ghanbarian and Sahimi (2017) demonstrated that the Ghanbarian et al. (2015a) model could accurately describe saturation-dependent electrical conductivity in mono-size packings of spheres, in accord with Einstein's relation that electrical conductivity is proportional to the diffusion coefficient.

**1.3. Objectives**

Universal scaling from percolation theory, in combination with scaling law from the effective-medium approximation, has been successfully used to describe lattice-Boltzmann simulations of gas and solute diffusion in overlapping and non-overlapping mono-sized sphere packs (Ghanbarian et al., 2015a). However, its application to experimental observations e.g., sand and glass bead packs has not been investigated yet. Furthermore, to the best of the authors' knowledge it has never been applied to describe the saturation dependence of gas permeability in porous media. Therefore, the main objective of this study is to apply concepts from percolation theory (PT) and the effective-medium approximation (EMA) to study gas diffusion and permeability experimentally in sand and glass bead packs under partially-saturated conditions. We compare PT, in combination with the EMA, to experimental measurements in packings of well-sorted angular and rounded sands as well as spherical glass beads.

**2. Theory**

PT and the EMA have been successfully applied to describe fluid flow and transport in lattices, pore networks and porous media (Hunt et al., 2014; Sahimi, 1994; 2011). Both approaches incorporate the effect of interconnectivity among pores in porous media, in contrast to bundle-of-tubes models in which each tube has no connectivity to others. One of the main features in PT and EMA is the presence of a percolation threshold below which fluid within the pore space loses its connectivity, and accordingly macroscopic transport coefficients e.g., gas diffusion and permeability vanish.

Following Ghanbarian and Hunt (2014) and Hunt et al. (2014) and using concepts from PT and the EMA, Ghanbarian et al. (2015a) proposed the following model to describe saturation-dependent transport coefficient $T(\varepsilon)$ in *porous media*:

$$\frac{T(\varepsilon)}{T(\phi)} = \begin{cases} \frac{\varepsilon_x - 2\phi/z}{\phi - 2\phi/z} \left[\frac{\varepsilon - \varepsilon_c}{\varepsilon_x - \varepsilon_c}\right]^2 & \varepsilon_c \leq \varepsilon \leq \varepsilon_x \\ \frac{\varepsilon - 2\phi/z}{\phi - 2\phi/z} & \varepsilon_x \leq \varepsilon \leq \phi \end{cases} \quad (3)$$

where the transport coefficient $T$ can represent diffusion coefficient or gas permeability, $\varepsilon_c$ is the critical gas-filled porosity for percolation below which there is no macroscopic flow or transport, and $\varepsilon_x$ is the crossover gas-filled porosity at which behavior transitions from percolation scaling to the EMA scaling. Eq. (3) was obtained by replacing $p$, $p_c$, $p_x$, and $2/z$ in Eq. (2) with $\varepsilon$, $\varepsilon_c$, $\varepsilon_x$, and $2\phi/z$ (for further detail, see Ghanbarian et al., 2015a). $z$ in Eq. (3) is the average pore coordination number. Coordination number represents the number of pore throats connected to the same pore body. Accordingly, in a porous medium there exists a distribution of coordination numbers, rather than a single unique value. For example, if the pore coordination number distribution follows a log-normal probability density function, $z$ would be the geometric mean value.

A special case of Eq. (3), i.e. $\varepsilon_x = \phi$, was successfully applied by Ghanbarian-Alavijeh and Hunt (2012b) to describe gas relative permeability in *natural porous media*. Regarding Eq. (3) and its applications, we should point out that Mu et al. (2007) also recommended that the saturation dependence of gas diffusion is nonlinear at low gas-filled porosities, while quasi-linear at high gas saturations. Furthermore, Eq. (3) has been successfully used to characterize lattice-Boltzmann simulations of saturation-dependent gas and solute diffusion in mono-size sphere packs (Ghanbarian et al., 2015a). For a recent review, see Hunt and Sahimi (2017). In the following, we accordingly compare

Eq. (3) with gas diffusion and gas permeability experimental measurements in sand and glass bead packs.

## 3. Materials and Methods

Experiments used in this study are from Hamamoto et al. (2016) in which angular sand (i.e., Granusil #30 and #50), rounded sand (i.e., Accusand #30/40 and #40/50), and glass bead (0.5 mm) were selected. Roundness and sphericity of sand grains were respectively 0.2 and 0.7 for Granusil and 0.8 and 0.8 for Accusand (see Table 1 from Hamamoto et al., 2016). Well-sorted sand grains (with diameter 0.3 or 0.5 mm) as well as spherical glass beads (with diameter 0.5 mm) were packed to reach a porosity of about 0.4. A total of five types of packs i.e., Granusil 0.3, Granusil 0.5, Accusand 0.3, Accusand 0.5, and Glass bead 0.5 were used to investigate gas transport in such porous media. Here, we briefly describe capillary pressure curve, gas diffusion, and gas permeability measurements as well as X-ray computed tomography data. Both capillary pressure curve and X-ray CT images provide some insights about microstructural properties of pores. The former, measured under drainage conditions, gives the pore-throat size distribution. The latter, however, can be used to capture pore-body and pore-throat size distributions as well as the average pore coordination number. For further detail, the interested reader is referred to the original published article by Hamamoto et al. (2016).

### 3.1. Capillary pressure curve

To measure capillary pressure curve, sands and/or glass beads were packed into stainless-steel cores with 2 cm height and an inner diameter of 5 cm to achieve the porosity of 0.4.

Water content was then measured at various capillary pressures e.g., $|P_c|$ = 0, 5, 10, 15, 20, 25, 30, 35, 40, 45, 50, 60, 70, and 90 cm $H_2O$ using the hanging water column method.

Although there exist several models in the literature, we apply the following fractal-like capillary pressure curve model to fit to the measured data (Ghanbarian et al., 2017):

$$\theta = \begin{cases} \phi - \beta \left[1 - \left(\frac{P_c}{P_d}\right)^{d_f-3}\right], & |P_d| \leq |P_c| \leq |P_{max}| \\ \phi, & |P_c| \leq |P_d| \end{cases} \quad (4)$$

where $\beta = \frac{(\phi-\theta_r)r_{max}^{3-d_f}}{r_{max}^{3-d_f}-r_{min}^{3-d_f}}$, $r_{min}$ and $r_{max}$ are respectively the smallest and largest pore throat radii, $P_c$ is the capillary pressure, $P_d$ is the displacement pressure, $\theta$ is the volumetric water content, $\phi$ is the porosity, $d_f$ is the pore space fractal dimension, and $\theta_r$ is the residual water content corresponding to the smallest pore throat or equivalently the maximum capillary pressure. The value of $\beta$ then can be determined via fitting Eq. (4) to measured capillary pressure curve. Eq. (4) was mathematically derived from a truncated power-law pore-throat size distribution and reduces to the Brooks-Corey model when $\beta = \phi - \theta_r$ and $\lambda = 3 - d_f$ (Ghanbarian et al., 2017).

Generally speaking, the greater the pore space fractal dimension, the broader the pore-throat size distribution. Although the pore space fractal dimension $d_f$ typically ranges between 2 and 3 in natural porous media, Ghanbarian-Alavijeh and Hunt (2012a) showed that one may expect $d_f$ to be a negative value, in accord with Mandelbrot (Mandelbrot, 1990). By directly fitting Eq. (4) to simulated capillary pressure curves of mono-size sphere packs, Ghanbarian and Sahimi (2017) reported -1.64 ≤ $d_f$ ≤ 0.62 and 0.99 ≤ $d_f$ ≤ 1.74 under drainage and imbibition conditions, respectively (see their Table 1).

## 3.2. Measurement of gas transport properties

Sand grains and glass beads were packed at pre-determined gravimetric water contents in 100-cm$^3$ stainless-steel cores and a bulk density corresponding to a porosity near 0.4. To reach a desired water content, water was added to pre-dried packings. Gas diffusion and permeability were measured under various water contents in repacked samples. Gas permeability was determined by flowing gas through the repacked samples at some flow rate between 0.3 and 5.0 L min$^{-1}$ and using Darcy's law. Diffusion coefficient was measured using the chamber method. Measurements were repeated three times in stainless-steel cores of 100 cm$^3$ (height of 5 cm and diameter of 5.1 cm). The interested reader is referred to the original publications by Hamamoto et al. (2016) for further information.

Equation (3) was directly fit to the diffusion and permeability data measured at various saturations in Excel. Parameters $\varepsilon_c$, $\varepsilon_x$, and $z$ were optimized by minimizing the square errors for each pack. The value of $z$ mainly depends on the slope of data at high gas-filled porosities, while $\varepsilon_x$ occurs when the slope switches from the linear to the nonlinear behavior.

## 3.3. X-ray computed tomography

An X-ray CT scanner was used for image analysis of the studied packings under completely dry conditions. Near 548 slices with resolution of 1024×1024 pixels were captured for each pack type. The average resolution of images was about 15 μm. Images were segmented via the edge-finding segmentation algorithm and by matching the porosity of the images and the measured one. To determine the average pore coordination number, the X-ray CT images were then analyzed using the 3DMA-Rock software package (Lindquist et al., 1996). The pore coordination number distribution in all packings followed the log-normal probability density function. Accordingly, we set $z$ as the geometric mean pore coordination number.

## 4. Results

In this section, we first present the results of fitting Eq. (4) to the measured capillary pressure curves. We then show how well Eq. (3) characterizes the saturation dependence of gas diffusion and gas permeability in sands and glass bead packs.

### 4.1. Capillary pressure curve

Using the Curve Fitting toolbox of MATLAB, we fit Eq. (4) to the measured drainage capillary pressure curves. We found that the pore space fractal dimension $d_f$ ranged between 0.982 (Glass bead 0.5) and 1.758 (Accusand 0.3). The fitted curves and the optimized capillary pressure curve model parameters are shown in Fig. 1 and reported in Table 1, respectively. As can be seen in Fig. 1, Eq. (4) fit the measured capillary pressure well ($R^2 > 0.95$; Table 1). We found no specific trend between the pore space fractal dimension and the grain size or shape in our sand packs. For example, the value of $d_f$ increased from 1.165 (Granusil 0.3) to 1.322 (Granusil 0.5) in Granusil sand packs, while decreased from 1.758 (Accusand 0.3) to 1.038 (Accusand 0.5) in Accusand packs. For Glass bead 0.5, we found $d_f = 0.982$, slightly greater than the maximum value reported for mono-size sphere packs by Ghanbarian and Sahimi (2017). In natural porous media, typically $2 < d_f < 3$. Therefore, $0.982 \leq d_f \leq 1.758$ obtained here shows that the pore-throat size distribution of the studied sand packs and glass beads is narrower than that in typical soils and rocks. In the Discussion section, we provide theoretical evidence that one should expect universal scaling e.g., Eq. (3) to describe gas diffusion and permeability in such packs with narrow pore-throat size distribution.

Generally speaking, the displacement pressure values for Granusil samples are greater than those for Accusand ones (see Table 1). We found $P_d$ = 13.5, 10.6, 9.7, and 9.4 cm $H_2O$ for Granusil 0.3, Granusil 0.5, Accusand 0.3, and Accusand 0.5, respectively. The values $P_d$ = 13.5 (Granusil 0.3) and 10.6 cm $H_2O$ (Granusil 0.5) are less than those reported by Sakaki and Illangasekare (2007) who determined the displacement pressure for similar sand packs and porosities. More specifically, Sakaki and Illangasekare (2007) found $P_d \approx$ 30.3 and 17.3 cm $H_2O$ for Granusil 0.3 and Granusil 0.5, respectively (see their Table 2). The discrepancies between 13.5 and 30.3 cm $H_2O$ for Granusil 0.3 as well as 10.6 and 17.3 cm $H_2O$ for Granusil 0.5 might be due to various sample dimensions used to measure the capillary pressure curve. In this study, samples dimensions are 2 cm (long) by 5 cm (inner diameter), while in their work 10 cm (long) by 8.25 cm (inner diameter). The effect of sample dimensions on capillary pressure curve, particularly near the saturation point, was addressed by Larson and Morrow (1981) and recently highlighted by Ghanbarian et al. (2015c).

**4.2. Gas diffusion and gas permeability**

*- Granusil 0.3*

Figure 2 shows measured gas diffusion (Fig. 2a) and gas permeability (Fig. 2b) as well as the fitted Eq. (3) for Granusil 0.3. As can be observed, the PT model (shown in blue) combined with the EMA (shown in red) fit the measured data well. For gas diffusion and permeability, we respectively found the critical gas-filled porosity $\varepsilon_c$ = 0.03 and 0.04, the crossover gas-filled porosity $\varepsilon_x$ = 0.23 and 0.17, and the average pore coordination number $z$ = 3.8 and 4.9 (Fig. 2). Interestingly, the critical gas-filled porosity values of the two gas transport mechanisms are very close (0.03 and 0.04). However, the crossover gas-filled porosity for gas diffusion (i.e., 0.23) was 35% greater than that for gas permeability. The average pore coordination number for gas diffusion ($z$ = 3.8) is near 22% less than that for gas permeability ($z$ = 4.9). Such a discrepancy most probably is because both gas and permeability were measured in different packs at various saturations. One should expect similar $z$ values from fitting Eq. (3) to saturation-dependent diffusion and permeability measured in the same pack.

Figure 2 clearly shows that both saturation-dependent gas diffusion and permeability follow a linear behavior at high gas-filled porosities, while conform to a nonlinear trend at low gas-filled porosity values. Linear trend in the $\varepsilon$-$k(\varepsilon)$ experiments has been previously observed see e.g., (Wang et al., 2014). However, the linear universal scaling from EMA has never been used before to describe linear saturation-dependent gas permeability.

- *Granusil 0.5*

Figure 3 presents results of fitting Eq. (3) to the measured gas diffusion and gas permeability for Granusil 0.5. Equation (3) fit the saturation-dependent gas diffusion experiment well, as shown in Fig. 3a. We found $\varepsilon_c$ = 0.04, $\varepsilon_x$ = 0.19, and $z$ = 4.5. However, the gas permeability data are scattered, which cause uncertainties in the optimized parameters $\varepsilon_c$ = 0.04, $\varepsilon_x$ = 0.16, and $z$ = 5.3 (see Fig. 3b). Similar to Granusil 0.3, we found identical critical gas-filled porosity for both gas diffusion and permeability. Likewise, we found that the greater the crossover gas-filled porosity, the smaller the average pore coordination number. Such a trend can be mathematically interpreted via an approximation proposed by Sahimi (1993), as we address in the following in section 5.1.

*- Accusand 0.3*

Results for Accusand 0.3 are shown in Fig. 4. The optimized parameters $\varepsilon_c$, $\varepsilon_x$, and $z$ are 0.03, 0.25, and 3.5 for gas diffusion and 0.04, 0.17, and 4.9 for gas permeability, respectively. Interestingly, these values are not greatly different from those obtained for Granusil 0.3 (compare Fig. 2 with Fig. 4). This means grain shape did not have a substantial impact on the Eq. (3) parameters. However, the effect of grain shape should be validated using further experimental data.

*- Accusand 0.5*

Figure 5 displays gas diffusion and gas permeability experiments for Accusand 0.5 as well as the fitted Eq. (3) to the measurements. As shown in Fig. 5a, $\varepsilon_c$ = 0.04, $\varepsilon_x$ = 0.25, and $z$ = 3.2 could well describe the measured gas diffusion data for Accusand 0.5. However, the gas permeability measurements are more scattered compared to the gas diffusion data (Fig. 5b vs. Fig. 5a). For gas permeability, we found $\varepsilon_c$ = 0.04, $\varepsilon_x$ = 0.17, and $z$ = 4.9. However, the optimized parameters are uncertain due to remarkable scatters in the measurements. Earlier, we stated that the scatters in the measurements might be because each data point was measured in a pack different than that used from other data points. This means although each sample was similarly repacked to reach the porosity of 0.4 or very close to that for all samples, the connectivity and microscopic pore space characteristics of the samples used to measure gas diffusion and gas permeability at various saturations are not necessarily identical.

*- Glass bead 0.5*

In addition to sand packs, we also compare the PT model, in combination with the EMA, in glass beads of average grain diameter of 0.5 mm (Glass beads 0.5). Results presented in Fig. 6 indicate that Eq. (3) describe the saturation dependence of gas diffusion and permeability in glass beads accurately. The model fit to both mechanisms well and similar to Figs. (2) to (5) the crossover gas-filled porosity for gas diffusion is greater than that for gas permeability. More specifically, we found $\varepsilon_c$ = 0.13 and 0.13, $\varepsilon_x$ = 0.30 and 0.27, and $z$ = 2.8 and 3 for gas diffusion and permeability, respectively (Fig. 6). We also found that $\varepsilon_c$ and $\varepsilon_x$ values are higher than those found in sand packs. However, these values are in agreement with $\varepsilon_c$ = 0.18, $\varepsilon_x$ = 0.22, and $z$ = 4.3 reported for a nonoverlapping sphere pack by Ghanbarian et al. (2015a).

In Fig. 7, we show gas diffusion coefficient as a function of gas-filled porosity for the five packs studied here. Due to scatters in measurements, the saturation dependence of permeability is not shown. As can be seen in Fig. 7, although all packs have porosity of 0.4, their saturation-dependent gas diffusion behaviors are different. At high gas-filled porosities, based on the EMA, Eq. (3) bottom line, the saturation dependence of gas diffusion is mainly controlled by the medium's connectivity (i.e., the average pore coordination number). However, at low gas-filled porosities, $D(\varepsilon)/D(\phi)$ depends remarkably on the critical gas-filled porosity. Since critical gas-filled porosity and average pore coordination number vary from one pack to another, one should expect different saturation-dependent gas diffusion in media studied here.

### 4.3. Estimation of coordination number from X-ray CT images

The average pore coordination number (the geometric mean) derived from X-ray CT images using the 3DMA-Rock software are presented in Table 2. We found $z = 3.10$, 3.07, 2.95, 3.23, and 3.79 for Granusil 0.3, Granusil 0.5, Accusand 0.3, Accusand 0.5, and Glass bead 0.5, respectively. These values are not greatly different from those reported by fitting Eq. (3) to the measured experiments, particularly for gas diffusion (see Table 2). One should note that the connectivity (the average coordination number) of two packs with the same porosity is not necessarily the same. Discrepancies between $z$ values from 3DMA-Rock and those obtained from fitting Eq. (3) to either gas diffusion or gas permeability measurements are most probably because X-ray images for each pack type (e.g., Accusand, Granusil, and Glass bead) were captured under completely dry conditions. In addition, as stated before, gas diffusion and permeability were measured at various saturations in different packs of the same porosity. This probably resulted in uncertainties in the calculation of $z$ value by fitting Eq. (3) to the measurements because the microstructure and topological properties of pore space in one pack might be different from another pack.

### 5. Discussion

Given that the number of experiments is limited and due to uncertainties in diffusion and permeability measurements at various saturations in different packs, it is not feasible to conclusively address the influence of grain size (0.3 and 0.5 mm) and shape (angular and rounded) in this study. The effect of grain shape and size on gas transport in partially saturated media, however, has been addressed in the literature. For example, Hamamoto et al. (2009) experimentally demonstrated that grain size substantially affects the effective diameter of pores contributing to gas transport near full saturation as well as at some intermediate saturation.

**5.1. Estimation of critical and crossover air-filled porosities**

In this section, we discuss methods that can be used to estimate critical and crossover air-filled porosity values. We also address that one should expect gas transport in the studied sand and glass bead packs should conform to universal scaling and Eq. (3).

*- Hunt (2004)*

As a first-order approximation, Hunt (2004) proposed that 10% of porosity provides a rough estimation of critical volumetric content for percolation ($\varepsilon_c = 0.1\phi$). Since the value of porosity of all packs is about 0.4 (Table 1), using Hunt's approach $\varepsilon_c$ should be about 0.04 (Table 3). Interestingly, this is in well agreement with those values reported for diffusion and permeability of gas in Granusil 0.3, Granusil 0.5, Accusand 0.3 and Accusand 0.5 (see Table 2). However, the value of 0.04 underestimates $\varepsilon_c$ for Glass bead 0.5 for which we found $\varepsilon_c = 0.13$.

*- Ghanbarian-Alavijeh and Hunt (2012b)*

Ghanbarian-Alavijeh and Hunt (2012b) found a crossover point $\theta_i$ in which the line slope of volumetric water content changes near saturated water content on capillary pressure curve (see their Fig. 6). At this crossover point, $\theta_i$, air starts percolating into the sand pack, in agreement with the results of Freijer (1994) who stated that, "… it can be concluded that the air-entry value gives a good estimate of the air-filled porosity (or the water content) at which pore blocking becomes relevant." Ghanbarian-Alavijeh and Hunt (2012b) demonstrated that one can have an estimate of the critical air-filled porosity by subtracting $\theta_i$ from the saturated water content i.e., $\varepsilon_c = \theta_s - \theta_i$ (see their Fig. 7). Following Ghanbarian-Alavijeh and Hunt (2012b), we estimated $\varepsilon_c$ from the capillary pressure curve measurements near the saturation point. Results obtained from sand packs as well as the Glass bead 0.5 are presented in Table 3. As can be seen, the Ghanbarian-Alavijeh and Hunt approach estimates $\varepsilon_c$ accurately not only for sand packs but also for the Glass bead 0.5 (see Table 3).

*- Percolation theory*

Bond percolation theory provides a theoretical method to estimate percolation threshold from bond coordination number. Using concepts from bond percolation, one can approximately determine the critical gas-filled porosity $\varepsilon_c$ in three-dimensional porous media as follows:

$$\varepsilon_c = \frac{3\phi}{2z} \tag{5}$$

Recall that $z$ is the average pore coordination number. The greater the $z$ value, the more connected the medium, and consequently the less the critical gas-filled porosity (or percolation threshold). This is in well agreement with the network theory results of Fatt (1960) on gas diffusion in partially-saturated porous media.

Sahimi (1993) stated that the region above but near the percolation threshold, where the universal quadratic power law (Eq. (3) top line) from percolation theory is valid, can be roughly estimated by

$$\varepsilon_x = \varepsilon_c + \frac{\phi}{z} \tag{6}$$

Eq. (6) indicates an inverse relationship between $\varepsilon_x$ and $z$. Ghanbarian et al. (2015a) demonstrated that Eqs. (5) and (6) estimated the critical and crossover gas-filled porosity values in overlapping and non-overlapping mono-sized sphere packs under perfectly wetting conditions accurately.

We approximated $\varepsilon_c$ and $\varepsilon_x$ via Eqs. (5) and (6), respectively. To estimate $\varepsilon_c$ for our five packs, we used the value of $z$ determined from X-ray CT images and the 3DMA-Rock software (see Table 2). The value of $\varepsilon_x$ was estimated from the same $z$ values as well as the $\varepsilon_c$ values determined using the Ghanbarian-Alavijeh and Hunt approach (see Table 3). Results, presented in Table 3, indicate that Eq. (5) remarkably overestimated $\varepsilon_c$ for sand packs (e.g., 0.19 vs. 0.03 for Granusil 0.3), while it accurately estimated $\varepsilon_c$ for the glass bead pack (0.16 vs. 0.13). We also found that the estimated $\varepsilon_x$ values given in Table 3 are in well agreement with those reported by fitting Eq. (3) to the measured saturation-dependent diffusion and permeability (Table 2), particularly for the later. Our results show that one can accurately estimate $\varepsilon_c$ and $\varepsilon_x$ values, if capillary pressure curve and X-ray CT images are available.

## 5.2. Validation of the universal scaling law

Using concepts from critical path analysis and universal scaling from percolation theory, Ewing and Hunt (2006) indicated that if

$$d_f \leq 3 - \frac{1}{2}\left[\frac{\phi - \varepsilon_c}{1 - \varepsilon_c}\right] \tag{7}$$

then transport coefficient (e.g., diffusion) should follow universal scaling such as Eq. (3). Using the $\varepsilon_c$ values determined via the Ghanbarian-Alavijeh and Hunt approach, we calculated the right side of Eq. (7) for each pack. Comparing results given in Table 3 with $d_f$ values presented in Table 1 indicate that one should expect the measured gas diffusion and permeability in sand and glass bead packs studied here to follow the universal scaling and Eq. (3). This means that the effect of the pore-throat size distribution on diffuion and permebaility of gas is minimal in our packs studied here.

As indicated here, saturation-dependent gas permeability was minimally affected by pore-throat size distribution. In contrast, saturation-dependent water permeability is expected to follow non-universal scaling in water-wet porous media (Hunt, 2001; Hunt e al., 2014). This is because of the fact that under partially saturated conditions, water is restricted to smallest pores, while gas exists in largest pores. Accordingly, the limiting behavior of pore-throat size distribution at zero pore size, which is known to be the origin of non-universality (Hunt e al., 2014), is not relevant to the saturation dependence of gas permeability. However, it could be relevant to that of water permeability. Generally speaking, one should expect Eq. (3) to accurately characterize nonwetting-phase relative permeability in porous media. Depending on the broadness of pore size distribution the crossover gas-filled porosity ($\varepsilon_x$) may occur somewhere between $\varepsilon_c$ and 1. For instance, in natural porous materials, such as soils and rocks Ghanbarian-Alavijeh and Hunt (2012b) demonstrated that $\varepsilon_x = \phi$. In well-sorted sand and glass bead packs, however, as we showed here, one should expect $\varepsilon_x < \phi$.

Eq. (3) is applicable to both wetting- and nonwetting-phase diffusion in porous materials. Experimental observations of Hunt et al. (2014) in natural porous media as well as lattice-Boltzmann simulation of Ghanbarian et al. (2015a) in mono-sized sphere packs indicated that Eq. (3) well described the saturation dependence of wetting- and nonwetting-phase diffusion coefficient.

In this study, we showed that concepts from percolation theory and the effective-medium approximation in combination with X-ray CT images and capillary pressure curve could accurately characterize saturation-dependent diffusion and permeability of gas in sand and glass bead packs. More specifically, we showed that both diffusion and permeability measurements conformed to the linear scaling from the effective-medium approximation at higher saturations, while to quadratic scaling from percolation theory at lower saturations, near and above the percolation threshold. Further investigations are required to study applications from percolation theory and the effective-medium approximation to gas transport in real soils and rocks using X-ray CT images and capillary pressure-saturation measurements. Our study was restricted to sand and glass bead packs with a porosity of 0.4. Further investigations are required to investigate the saturation dependence of gas diffusion and permeability in packings of different porosities.

## 6. Conclusions

In this study, we evaluated a theoretical model from percolation theory and the effective-medium approximation using gas diffusion and gas permeability experiments measured in sand packs composed of either angular or rounded grains of near the same size. We found that both gas diffusion and gas permeability in such media showed almost linear behavior at high saturations (high gas-filled porosities), while below some crossover point it switched to a nonlinear trend near a critical gas-filled porosity. Comparing theory with the experiments showed that the universal quadratic scaling law from percolation theory, combined with the universal linear scaling law from the EMA, provided an excellent description of saturation-dependent gas diffusion and gas permeability over the entire range of saturation in our packs. Because the number of experiments was limited, and both diffusion and permeability were measured at various saturations in different packs, it was not feasible to address the effect of grain size and shape in this study conclusively.

**Acknowledgement**

BG is grateful to Kansas State University for supports through faculty startup funds. The sand samples were originally provided by the Center for Experimental Study of Subsurface Environmental Processes, Colorado School of Mines, Golden CO, USA.

**Figure captions**

Fig. 1. Measured capillary pressure curves for five samples used in this study. The curves represent the fitted Eq. (4) to the measurements. Fitted parameters for each sample are presented in Table 1.

Fig. 2. Gas diffusion (a) and gas permeability (b) as a function of gas-filled porosity for Granusil 0.3. The blue and red lines represent respectively percolation theory (PT) and the effective-medium approximation (EMA) scaling laws, Eq. (3), top line and bottom line.

Fig. 3. Gas diffusion (a) and gas permeability (b) as a function of gas-filled porosity for Granusil 0.5. The blue and red lines represent respectively percolation theory (PT) and the effective-medium approximation (EMA) scaling laws, Eq. (3), top line and bottom line.

Fig. 4. Gas diffusion (a) and gas permeability (b) as a function of gas-filled porosity for Accusand 0.3. The blue and red lines represent respectively percolation theory (PT) and the effective-medium approximation (EMA) scaling laws, Eq. (3), top line and bottom line.

Fig. 5. Gas diffusion (a) and gas permeability (b) as a function of gas-filled porosity for Accusand 0.5. The blue and red lines represent respectively percolation theory (PT) and the effective-medium approximation (EMA) scaling laws, Eq. (3), top line and bottom line.

Fig. 6. Gas diffusion (a) and gas permeability (b) as a function of gas-filled porosity for Glass bead 0.5. The blue and red lines represent respectively percolation theory (PT)

and the effective-medium approximation (EMA) scaling laws, Eq. (3), top line and bottom line.

Fig. 7. Gas diffusion as a function of gas-filled porosity for all five packs studied here. The blue and red lines represent respectively percolation theory (PT) and the effective-medium approximation (EMA) scaling laws, Eq. (3), top line and bottom line. Parameters of Eq. (3) for each pack are given in Table 2.

Table 1. Salient properties of the samples used in this study. $\beta$, $d_f$, $P_d$, and $R^2$ were determined by directly fitting Eq. (4) to the capillary pressure curve measurements.

| Sand pack | $\rho_b$ (gr cm$^{-3}$) | $\rho_s$ (gr cm$^{-3}$) | $\phi$ (cm$^3$ cm$^{-3}$) | $\beta$ | $d_f$ | $P_d$ (cm) | $R^2$ |
|---|---|---|---|---|---|---|---|
| Granusil 0.3 | 1.57 | 2.65 | 0.41 | 0.414 | 1.165 | 13.5 | 0.99 |
| Granusil 0.5 | 1.57 | 2.65 | 0.41 | 0.418 | 1.322 | 10.6 | 0.99 |
| Accusand 0.3 | 1.60 | 2.66 | 0.40 | 0.437 | 1.758 | 9.7 | 0.95 |
| Accusand 0.5 | 1.59 | 2.66 | 0.40 | 0.407 | 1.038 | 9.4 | 0.97 |
| Glass bead 0.5 | 1.57 | 2.62 | 0.40 | 0.400 | 0.982 | 8.9 | 0.98 |

$\rho_b$ is bulk density, $\rho_s$ is particle density, $\phi$ is porosity, $\beta$ is capillary pressure curve model parameter, $d_f$ is pore space fractal dimension, $P_d$ is displacement pressure, $R^2$ is correlation coefficient.

Table 2. Critical and crossover gas-filled porosity values as well as average pore coordination number for various packings studied here.

| Pack | Gas diffusion Eq. (3) | | | Gas permeability Eq. (3) | | | 3DMA-Rock |
|---|---|---|---|---|---|---|---|
| | $\varepsilon_c$ | $\varepsilon_x$ | $z$ | $\varepsilon_c$ | $\varepsilon_x$ | $z$ | $z^*$ |
| Granusil 0.3 | 0.03 | 0.23 | 3.8 | 0.04 | 0.17 | 4.9 | 3.10 (0.76) |
| Granusil 0.5 | 0.04 | 0.19 | 4.5 | 0.04 | 0.16 | 5.3 | 3.07 (0.74) |
| Accusand 0.3 | 0.03 | 0.25 | 3.5 | 0.04 | 0.17 | 4.9 | 2.95 (0.74) |
| Accusand 0.5 | 0.04 | 0.25 | 3.2 | 0.04 | 0.17 | 4.9 | 3.23 (0.94) |
| Glass bead 0.5 | 0.13 | 0.30 | 2.8 | 0.13 | 0.27 | 3.0 | 3.79 (1.27) |

* The geometric mean pore coordination number derived from X-ray CT images and its variance in parentheses.

Table 3. Estimated critical and crossover gas-filled porosities using different methods for various packs studied here.

| Pack | Hunt (2004) $\varepsilon_c = 0.1\phi$ | Ghanbarian-Alavijeh and Hunt (2012b) $\theta_i$ | Ghanbarian-Alavijeh and Hunt (2012b) $\varepsilon_c (= \theta_s - \theta_i)$ | Eq. (5) $\varepsilon_c^*$ | Eq. (6) $\varepsilon_x^{**}$ | $3 - \frac{1}{2}\left[\frac{\phi-\varepsilon_c}{1-\varepsilon_c}\right]$ † |
|---|---|---|---|---|---|---|
| Granusil 0.3 | 0.04 | 0.38 | 0.03 | 0.19 | 0.16 | 2.804 |
| Granusil 0.5 | 0.04 | 0.38 | 0.03 | 0.20 | 0.16 | 2.804 |
| Accusand 0.3 | 0.04 | 0.38 | 0.02 | 0.20 | 0.16 | 2.806 |
| Accusand 0.5 | 0.04 | 0.34 | 0.06 | 0.19 | 0.18 | 2.819 |
| Glass bead 0.5 | 0.04 | 0.29 | 0.11 | 0.16 | 0.22 | 2.837 |

* $z$ was obtained via 3DMA-Rock from X-ray CT images (see Table 2).
** $z$ was determined via 3DMA-Rock (see Table 2) and $\varepsilon_c$ was estimated using the Ghanbarian-Alavijeh and Hunt (2012b) approach.
† $\varepsilon_c$ was estimated using the Ghanbarian-Alavijeh and Hunt (2012b) approach.

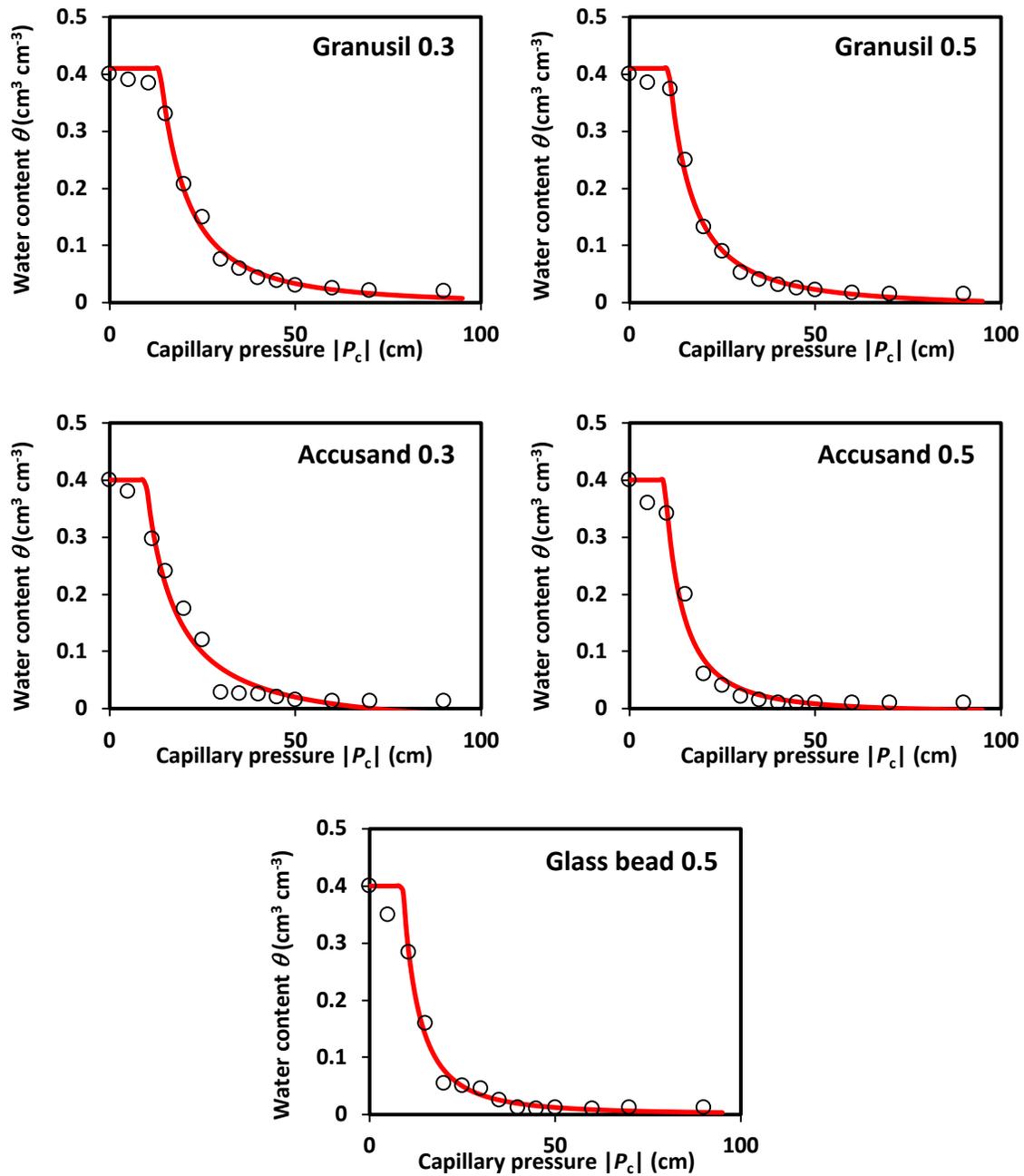

Fig. 1. Measured capillary pressure curves for five samples used in this study. The curves represent the fitted Eq. (4) to the measurements. Fitted parameters for each sample are presented in Table 1.

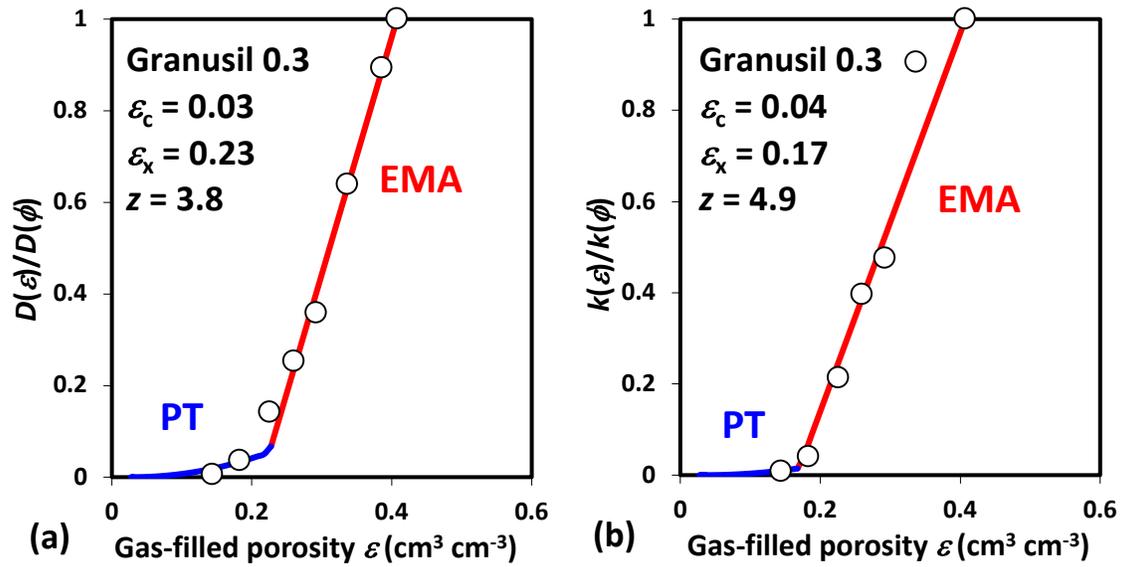

Fig. 2. Gas diffusion (a) and gas permeability (b) as a function of gas-filled porosity for Granusil 0.3. The blue and red lines represent respectively percolation theory (PT) and the effective-medium approximation (EMA) scaling laws, Eq. (3), top line and bottom line.

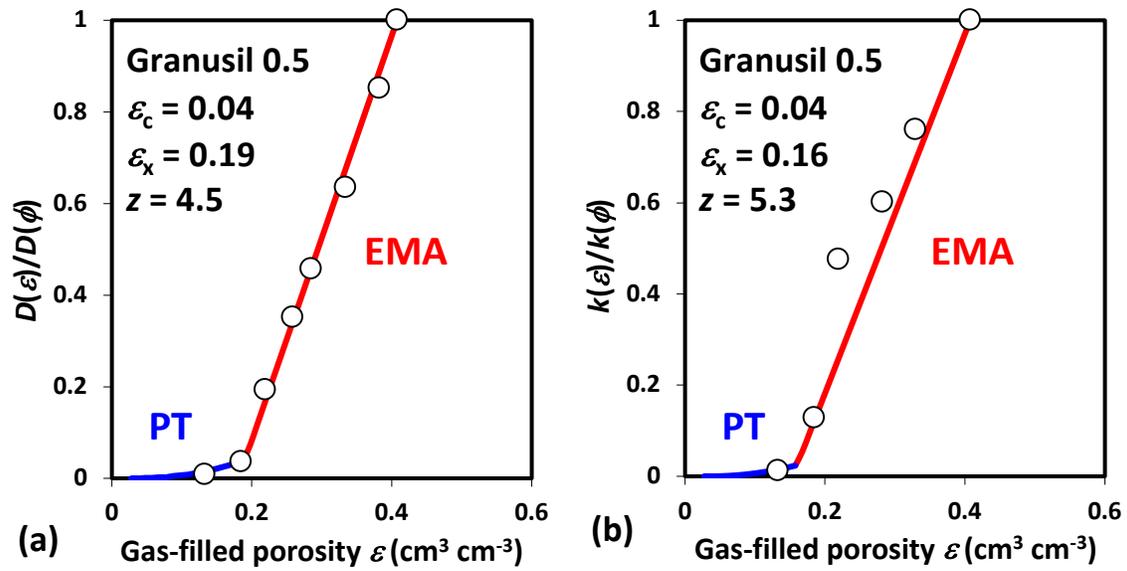

Fig. 3. Gas diffusion (a) and gas permeability (b) as a function of gas-filled porosity for Granusil 0.5. The blue and red lines represent respectively percolation theory (PT) and the effective-medium approximation (EMA) scaling laws, Eq. (3), top line and bottom line.

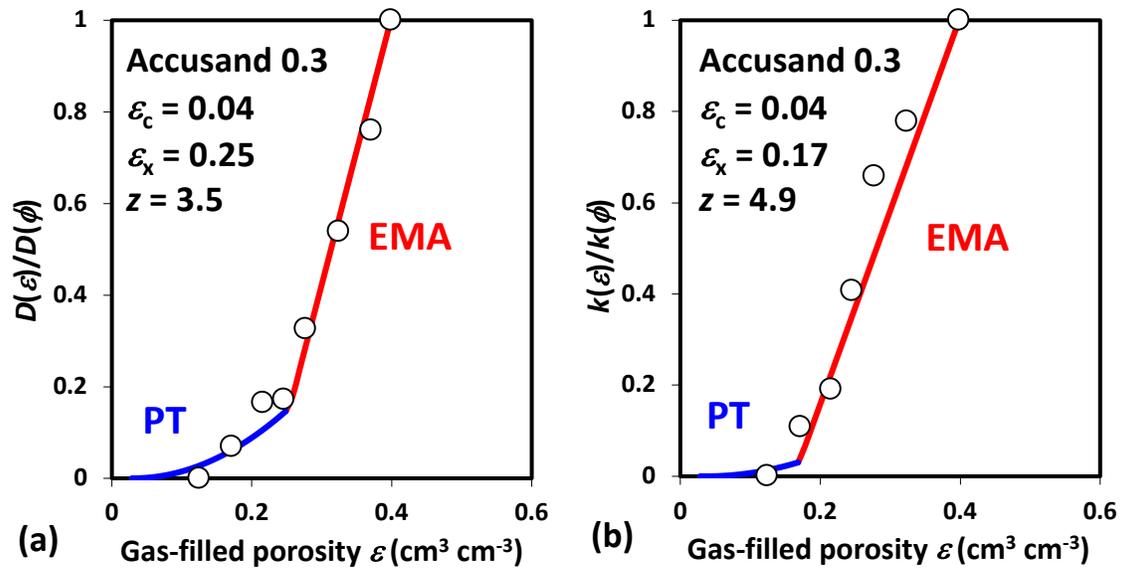

Fig. 4. Gas diffusion (a) and gas permeability (b) as a function of gas-filled porosity for Accusand 0.3. The blue and red lines represent respectively percolation theory (PT) and the effective-medium approximation (EMA) scaling laws, Eq. (3), top line and bottom line.

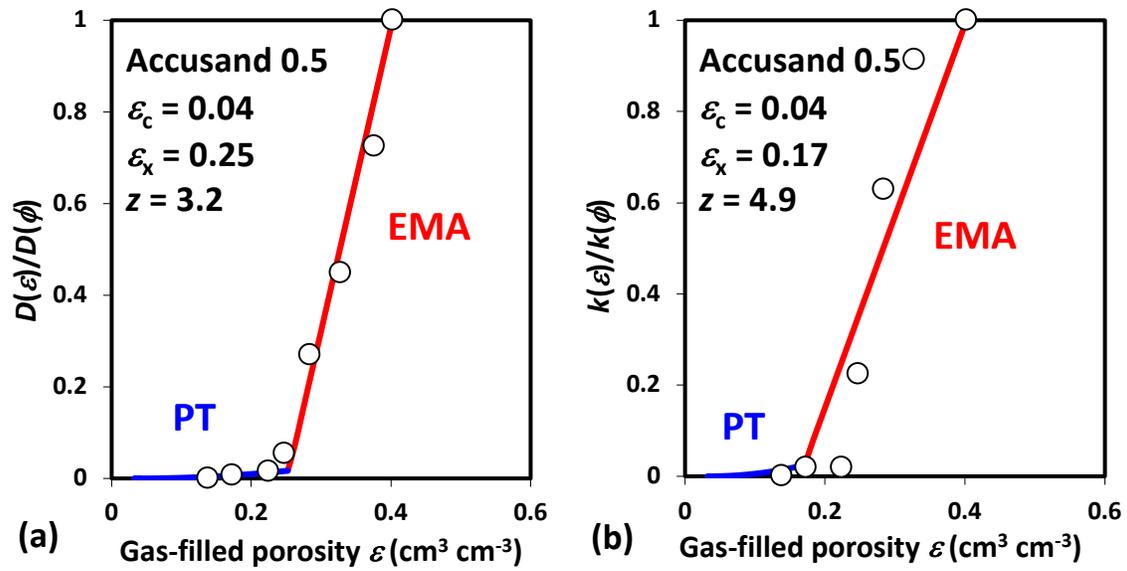

Fig. 5. Gas diffusion (a) and gas permeability (b) as a function of gas-filled porosity for Accusand 0.5. The blue and red lines represent respectively percolation theory (PT) and the effective-medium approximation (EMA) scaling laws, Eq. (3), top line and bottom line.

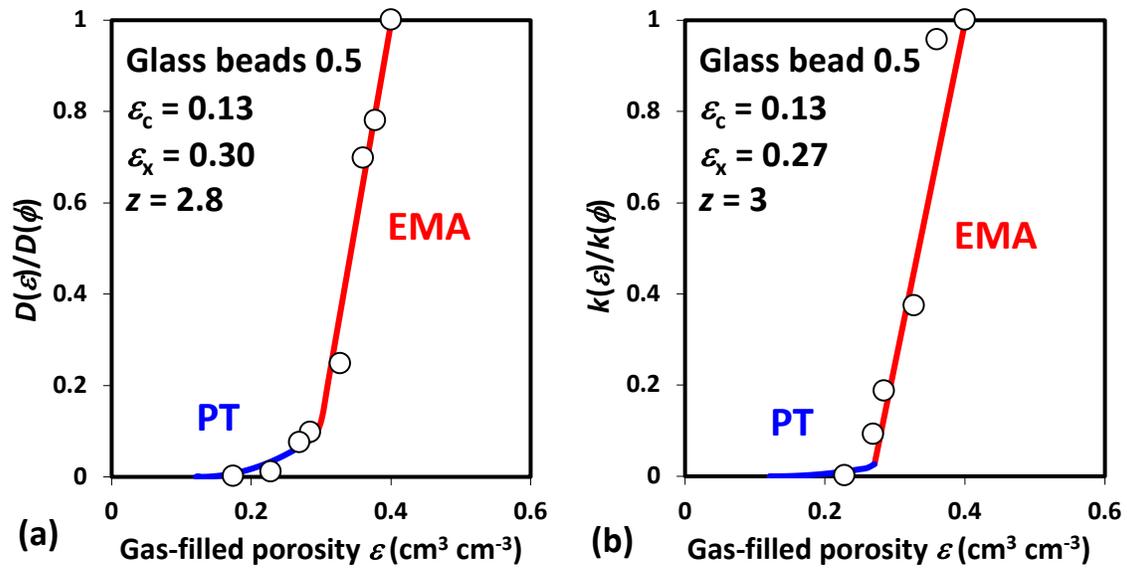

Fig. 6. Gas diffusion (a) and gas permeability (b) as a function of gas-filled porosity for Glass bead 0.5. The blue and red lines represent respectively percolation theory (PT) and the effective-medium approximation (EMA) scaling laws, Eq. (3), top line and bottom line.

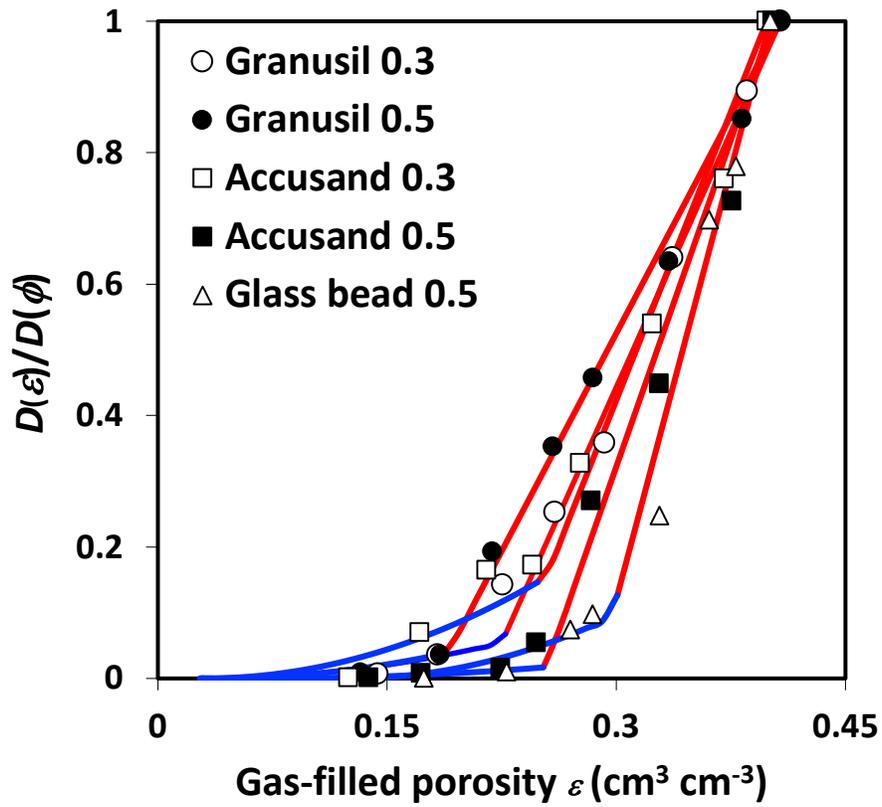

Fig. 7. Gas diffusion as a function of gas-filled porosity for all five packs studied here. The blue and red lines represent respectively percolation theory (PT) and the effective-medium approximation (EMA) scaling laws, Eq. (3), top line and bottom line. Parameters of Eq. (3) for each pack are given in Table 2.